# HIGH BRIGHTNESS AND FULLY COHERENT X-RAY PULSES FROM XFELO SEEDED HIGH-GAIN FEL SCHEMES *

H. X. Deng[#], C. Feng,  SINAP/SSRF, Shanghai, 201800, China

*Abstract*

   The successful operation of the hard x-ray self-seeding experiment at the LCLS opens the era of fully coherent hard x-ray free electron lasers (FELs). However, the shot-to-shot radiation fluctuation is still a serious issue. In this paper, high-gain, single-pass x-ray FEL schemes seeded by the narrow bandwidth radiation signal from an x-ray FEL oscillator were proposed and investigated, which are expected to generate high brightness, fully coherent and stable x-ray pulse. A simple model has been developed to figure out the temporal and the spectral structures of the output pulses in x-ray FEL oscillator. And options using two synchronized accelerators and using one accelerator were considered, respectively.

## INTRODUCTION

   With the great successful of LCLS [1] and SACLA [2], the first hard x-ray FEL and the shortest wavelength FEL in the world, many hard x-ray FEL projects, driven by the growing interests of FEL users, are being constructed and planned around the world. Currently, all the hard x-ray FELs use self-amplified spontaneous emission (SASE) [3] as the lasing mode, which starts from the electron beam noise, and results in radiation with transverse coherence, but poor longitudinal coherence. In order to generate fully coherent radiation, various seeded FELs were proposed and intensively studied worldwide [4-8]. Echo-enabled harmonic generation (EEHG) [9] is one of the frequency up-conversion schemes, which could efficiently work at several tens of harmonic of a commercial seed laser. The experimental demonstration of the EEHG mechanism [10] and the first lasing of an EEHG-FEL [11] pave the way to coherent soft x-ray radiation.

   In the hard x-ray regime, the self-seeding approach was proposed to narrow the bandwidth of SASE radiation, and has been experimentally demonstrated at LCLS [12]. In a self-seeding scheme, the noisy SASE radiation generated in the first undulator section is spectrally purified by a crystal filter. Then, in the second undulator section, this spectrally purified FEL pulse serves as a highly coherent seed to interact with the electron bunch again to configure a FEL amplifier, which could significantly improve the temporal coherence of the final output radiation. However, self-seeding scheme suffers from the serious shot-to-shot fluctuations. With the development of high-reflectivity high-resolution x-ray crystal and ultra-low emittance ERL beam, the low-gain oscillator was reconsidered as one of the candidates for hard x-ray FEL, known as XFELO [13-16]. It is widely believed that XFELO could provide fully coherent and stable x-ray pulses. Therefore, in order to obtain the high peak brightness, fully coherent and stable x-ray radiation pulses while avoiding the  the heat loading and instability issues of the x-ray crystal cavity, the high-gain, single-pass x-ray FEL configurations driven by the spectrally pure seed from XFELO were proposed in this paper. We tried to establish a straightforward model for effectively resolving the longitudinal mode of XFELO. With the help of this model, options using the electron beam from two different accelerators and from the same accelerator were studied, respectively.

## LONGITUDINAL MODE OF XFELO

   As the schematic of XFELO illustrated in Fig. 1, when the electron bunch with current distribution of $I(t)$ passes through the undulator, the co-propagating radiation pulse $P_{cav}(t)$ will be enhanced by a factor of $G(t)$. Usually, the x-ray cavity was established by a couple of Bragg crystals with reflectivity of $R(\omega)$, which largely depends on the photon energy and the Bragg energy of crystal. Then one may have the radiation power in the cavity

$$P_{cav}^{n+1}(t) = P_{cav}^n(t)[1 + G^n(t)] * R^2(t). \quad (1)$$

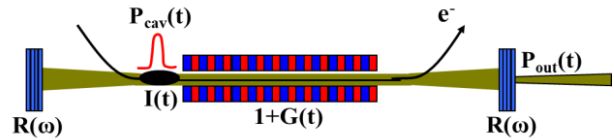

Figure 1: Schematic and brief model of an XFELO.

$G^n(t)$ is the small-signal gain of the radiation pulse in the *n*-th roundtrip. Without considering its dependence on the radiation intensity, for a Gaussian beam current with pulse duration of $\sigma$ and the current distribution of $I(t)$, the gain can be given as the exponential decay form of the maximum gain $G_0$, as follows,

$$G^n(t) = G_0 e^{-\frac{t^2}{2\sigma^2}} \propto I(t). \quad (2)$$

$R(t)$ is the total reflectivity of the x-ray cavity. As a rough estimate, we can write it as

$$R(t) = F^{-1}(R) = R_0 \frac{1}{2\pi} \int H(\frac{\omega-\omega_c}{\Delta}) e^{-i\omega t} d\omega, \quad (3)$$

where $R_0$, $\Delta$, $\omega_c$ and $H$ is the crystal reflectivity for the radiation pulse of the critical frequency, the bandwidth and the critical frequency of Bragg crystal, and the rectangular function respectively. And then the output radiation pulse can be simply presented as

___________________
*Work supported by Major State Basic Research Development Program of China (2011CB808300) and the National Natural Science Foundation of China (11175240 and 11205234).

#denghaixiao@sinap.ac.cn

$$P_{out}^{n+1}(t) = P_{cav}^n(t)[1 + G^n(t)][1 - R(t)], \quad (4)$$

Since the small-signal gain is larger than the round-trip cavity loss, the radiation evolves from initial spontaneous emission to a coherent pulse. After an exponential growth, the radiation saturates due to the gain decrease caused by over-modulation of the high intensity radiation in XFELO cavity. The XFELO principle is very similar with a laser oscillator, and the transverse mode at XFELO saturation can be resolved from an ABCD matrix analysis. However, serving as a seed laser of high-gain, single-pass FEL, the longitudinal mode of XFELO is of much more interest.

In principle, the 3D numerical simulation is capable for calculating the XFELO modes, but it is time-consuming and uneconomic for preliminary optimization. In order to efficiently figure out the longitudinal mode of the XFELO, Equations (1)-(4) are numerically resolved with an initial Gaussian radiation with arbitrary intensity. Table 1 gives three groups of parameters we used for simulation, where the gain bandwidth is less, comparable and larger than the Bragg bandwidth of the crystal mirrors.

Table 1: Main parameters used for longitudinal mode calculation of an XFELO

| Gain bandwidth | σ | $G_0$ | $R_0$ | Δ |
|---|---|---|---|---|
| 15meV | 125fs | ~ 0.56 | 80% | 30meV |
| 74meV | 25fs | ~ 1.61 | 80% | 30meV |
| 186meV | 10fs | ~ 4.56 | 80% | 30meV |

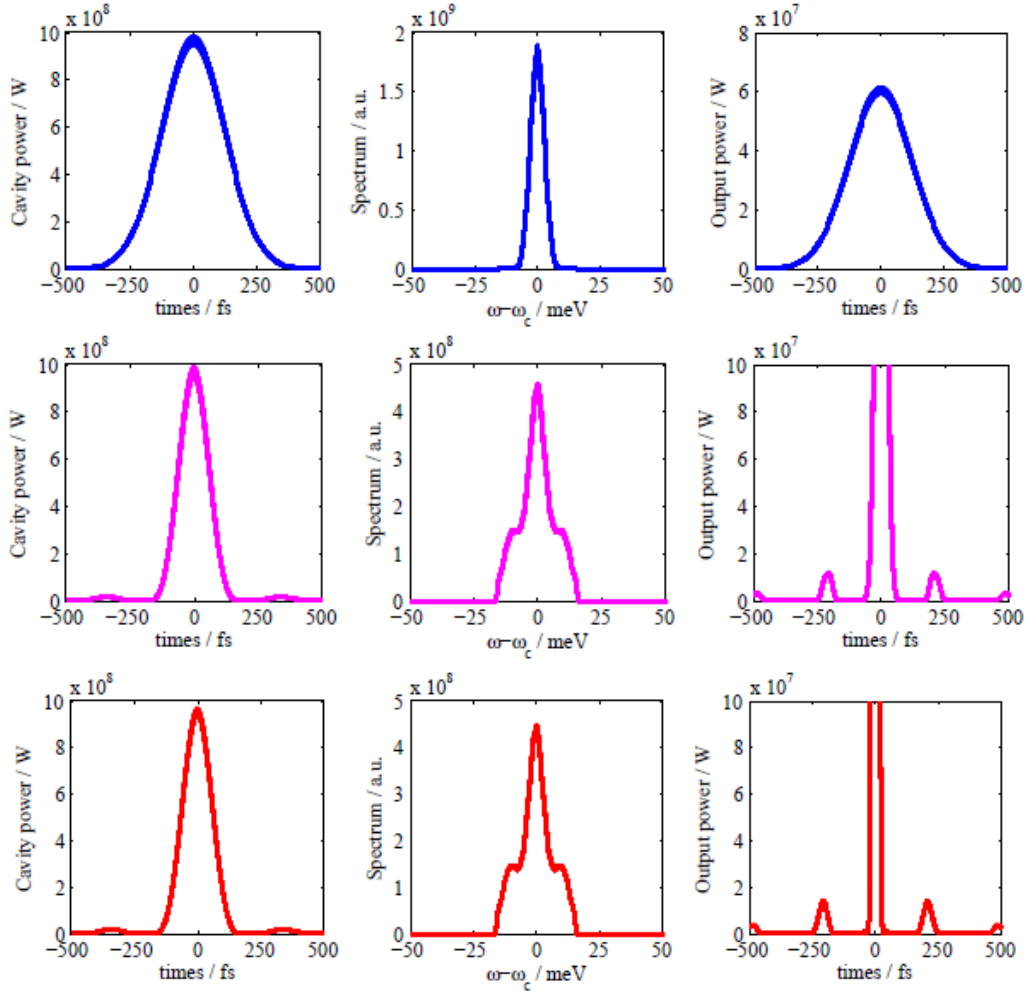

Figure 2: The longitudinal mode of XFELO, where the blue, pink and red illustrate the cases with 125fs, 25fs and 10fs electron bunch respectively. Because the gain reduction caused by the increasing radiation power was not included, the gain $G_0$ here should be the single-pass gain when achieving the XFELO saturation. Under such circumstance, one may find a steady longitudinal mode of XFELO. Since the radiation power level can be controlled by the cavity detuning and by varying the electron beam parameters, we assume to maintain a 1GW radiation power in the cavity.

For a rather long electron bunch, as shown in Fig. 2, the gain bandwidth is small, and the crystal will not affect the radiation pulse much once the coherence was built up in the cavity. Then the longitudinal mode is determined by the electron bunch as that of in infrared FEL oscillators. For a short electron bunch, as the 10fs example shown in Fig. 2, the longitudinal mode is no more related to the electron bunch length. Since the high resolution of Bragg crystal, the radiation bandwidth at the undulator entrance is always confined within 30meV and it results a 60fs pulse duration. Since the gain bandwidth is 184meV, the undulator naturally produces the radiation components out of 30meV, which will be filtered immediately at the downstream crystal. Then the downstream crystal works exactly as the crystal in self-seeding configuration [12], and a wake radiation pulse can be expected in the XFELO output pulse. Indeed, a 10MW peak power wake pulse is found 200fs behind the main pulse in the calculations.

## XFELO SEEDED HIGH-GAIN FEL SCHEMES

We first consider the XFELO seeded high-gain x-ray FEL schemes with two synchronized accelerators, e.g., an ERL or ultimate storage ring based XFELO is used as a seed of high-gain FEL driven by a state-of-the-art LINAC [17]. In this configuration, two accelerators are needed to be synchronized within 10fs accuracy, when considering the bunch length of high-brightness LINAC. According to FEL physics, the XFELO seed will dramatically reduce the x-ray FEL bandwidth, i.e., from the pierce parameter order to tens of meV level, and thus enhance the x-ray brightness by 3~4 orders of magnitude.

Now we discuss the possibility of XFELO seeded high-gain FELs with a high-repetition superconductive LINAC such as European-XFEL. A Silicon based XFELO for the European-XFEL was proposed [18], intensive simulations show that the single-pass gain larger than 300% can be achieved by using 3 segments SASE1 undulator with the electron beam of 0.5nC charge and 50fs bunch length. Considering the thermal loading effects of the Megahertz repetition pulse energy as high as tens of milliJoule stored in the x-ray cavity, here we try to propose high-gain FEL configuration seeded by the XFELO for SASE1 beam line, as illustrated in Fig. 3.

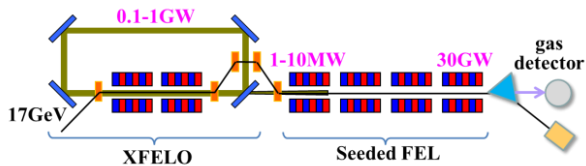

Figure 3: XFELO seeded high-gain FEL configuration.

For the long electron bunch, the XFELO output pulse can be directly served as the seed for the following single-pass FEL. When the electron bunch becomes short, the temporal structure of XFELO is obviously affected by the Bragg crystal. Thus if the LINAC was operated in low-charge mode, e.g., the 20pC case, the crystal downstream of the XFELO undulator results in a wake radiation pulse. If the peak power in the x-ray cavity is maintained at 0.1-1GW level, the wake pulse power is MW order, which can be used as the seed for the followed high-gain part. Since the interval between the main pulse and the wake pulse is about 200fs, a magnetic chicane with dispersion of 120μm is needed for electron beam delay. The 1MW seed can be amplified to 30GW in 100m further undulator, as given in Fig. 4. Moreover, the bunch to bunch energy deviations for the European-XFEL in the order of $10^{-4}$ do not influence the stored energy of XFELO significantly, thus the output radiation fluctuation of the seeded FEL is expected to be small and can be stably amplified to TW level with an undulator tapering technique [19].

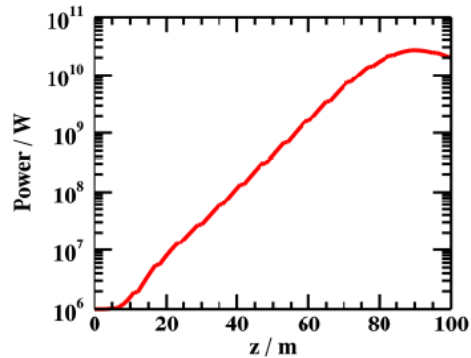

Figure 4: Power growth in XFELO seeded high-gain FEL.

## CONCLUSIONS & OUTLOOK

The XFELO seeded high-gain x-ray FEL schemes were proposed for generating fully coherent and stable x-ray radiation pulses. The simple XFELO longitudinal model predicts a wake radiation pulse as in self-seeding scheme, which can be used as the seed in further amplification. However, it is worth to stress that the analytical model is not perfect currently, and this work is still preliminary. Therefore, a more sophisticated model with the transverse effects, the gain reduction on the beam energy spread and the cavity length detuning is under development by using MATLAB SIMULINK toolkits. And the 3D start-to-end simulations will also be carried out in future.

## ACKNOWLEDGMENT

The authors are grateful to D. Z. Huang and C. K. Yin for helpful discussions.